\documentclass[preprint,number,12pt]{elsarticle}

\usepackage[makeroom]{cancel}

\usepackage{upgreek} 

\usepackage[utf8]{inputenc}
\usepackage[T1]{fontenc}
\usepackage{qcircuit}

\usepackage{float}
\usepackage[normalem]{ulem}
\usepackage{graphicx, xcolor, graphpap}
\usepackage{enumitem}
\usepackage{amssymb}
\usepackage{amsthm}
\usepackage{multirow}
\usepackage[colorlinks=true,citecolor=blue,linkcolor=magenta]{hyperref}
\usepackage[T1]{fontenc}
\usepackage{verbatim}
\usepackage{mathtools}
\usepackage{titlesec}
\usepackage{natbib}

\long\def\ca#1\cb{} 

\newcommand{\abs}[2][]{#1| #2 #1|}

\newcommand{\bramatket}[3]{\langle #1 \hspace{1pt} | #2 | \hspace{1pt} #3 \rangle}

\newcommand{\avg}[1]{\langle #1\rangle }
\newcommand{\dya}[1]{\ket{#1}\!\bra{#1}}
\global\long\def\re{\mathrm{Re}}
\global\long\def\im{\mathrm{Im}}

\newcommand{\ave}[1]{\langle #1\rangle}               
\renewcommand{\geq}{\geqslant}

\newcommand{\ad}{^\dagger}

\newcommand{\eq}{\text{eq}}

\newcommand{\bra}[1]{\left\langle #1\right|}
\newcommand{\ket}[1]{\left|#1\right\rangle}

\newcommand{\tr}[1]{\mathrm{tr}\left\{#1\right\}}
\newcommand{\ptr}[2]{\mathrm{tr_{#1}}\left\{#2\right\}}

\newcommand{\e}[1]{\exp{\left(#1\right)}}

\newcommand{\id}{\mathbb{I}}

\newcommand{\bla}{bla\\bla\\bla\\bla\\bla}
\newcommand{\mb}[1]{\mbox{\boldmath$#1$}}
\newcommand{\mc}[1]{\mathcal{#1}}

\newcommand{\mf}[1]{\mathfrak{#1}}
\newcommand{\mrm}[1]{\mathrm{#1}}

\global\long\def\eqn#1{\begin{align}#1\end{align}}

\global\long\def\ket#1{\left|#1\right\rangle }
\global\long\def\bra#1{\left\langle #1\right|}
\global\long\def\bkt#1{\left(#1\right)}
\global\long\def\sbkt#1{\left[#1\right]}
\global\long\def\cbkt#1{\left\{#1\right\}}
\global\long\def\abs#1{\left\vert#1\right\vert}

\global\long\def\der#1#2{\frac{{d}#1}{{d}#2}}

\global\long\def\dd{\mathrm{d}}

\global\long\def\avg#1{\left\langle #1 \right\rangle}
\global\long\def\mr#1{\mathrm{#1}}
\global\long\def\mb#1{{\mathbf #1}}
\global\long\def\mc#1{\mathcal{#1}}

\global\long\def\non{\nonumber}

\newcommand{\overbar}[1]{\mkern 1.5mu\overline{\mkern-1.5mu#1\mkern-1.5mu}\mkern 1.5mu}
\newcommand\dbar[1]{{%
  \setbox0=\hbox{$\overbar{#1}$}%
  \ht0=\dimexpr\ht0-.15ex\relax%
  \overbar{\copy0}%
}}

{}
{}

\journal{Encyclopedia of Mathematical Physics 2nd edition}

\begin{document}

\begin{frontmatter}

\title{Thermodynamic perspective on quantum fluctuations}

\author[inst1]{Akira Sone}
\affiliation[inst1]{{Department of Physics, University of Massachusetts, Boston,},
            city={Boston},
            postcode={MA 02125}, 
            country={USA}}
\ead{akira.sone@umb.edu}
\author[inst2]{Kanu Sinha}
\affiliation[inst2]{{Wyant College of Optical Sciences, University of Arizona},
            city={Tucson},
            postcode={AZ 85721}, 
            country={USA}}
\ead{kanu@arizona.edu}
\author[inst3]{Sebastian Deffner}
\affiliation[inst3]{{Department of Physics, University of Maryland, Baltimore County,},
            city={Baltimore},
            postcode={MD 21250}, 
            country={USA}}
\ead{deffner@umbc.edu}

\begin{abstract}
What is the major difference between large and small systems? At small length-scales the dynamics is dominated by fluctuations, whereas at large scales fluctuations are irrelevant. Therefore, any thermodynamically consistent description of quantum systems necessitates a thorough understanding of the nature and consequences of fluctuations. In this chapter, we outline two closely related fields of research that are commonly considered separately -- fluctuation forces and fluctuation theorems. Focusing on the main gist of these exciting and vivid fields of modern research, we seek to provide a instructive entry point for both communities of researchers interested in learning about the other.
\end{abstract}




\end{frontmatter}


\section{Introduction}

Thermodynamics is one of the, if not the most powerful theory in theoretical physics. Built upon macroscopic observation and empirical laws, thermodynamics statements remain true -- independent of the microscopic details of the system under study. Therefore, to apply and utilize thermodynamics it is always instrumental to identify the good macroscopic observables, which are measurable quantities immune to fluctuations in time and space, over the time- and length-scales of external observation \cite{Callen1998}.

Focusing on macroscopic variables and quasi-static processes, thermodynamics permits a tremendous reduction of mathematical complexity in the description of many-body systems. However, all real-life processes occur at finite-rates and it is frequently desirable to describe also the microscopic details of fluctuations around the macroscopically observable mean values. This becomes particularly important for quantum systems, for which fluctuations are comprised of both, thermal noise and quantum uncertainty \cite{DeffnerBook19}.

Curiously, the study of fluctuations in quantum systems has been developed in (at least) two very different directions, that are commonly presented without much overlap. Quantum stochastic thermodynamics \cite{DeffnerBook19,Strasberg2022,Myers2022AVSQS} focuses on formulations of the laws of thermodynamics for quantum systems, whereas Casimir physics \cite{Milonni, DalvitMilonni,  Buhmann1, Buhmann2, ForcesQV15} studies the mechanically measurable consequences of fluctuations of quantum electromagnetic (EM) fields.

Since this two directions of research are obviously closely related, or at least they should be conceptually close, it appears desirable to have a coherent presentation. In this entry to the \emph{Encyclopedia of Mathematical Physics} we provide, therefore, a concise outline of the main notions, which may serve as an entry point into the large body of literature on the two topics.

\section{Fluctuations in quantum fields and their corresponding forces}

In Quantum Field Theory (QFT), empty space is not quite empty but jostling with fluctuations, both thermal and quantum in origin. Quantum fluctuations are an  inevitable and fascinating feature of QFT, persisting in the absence of all matter and radiation, even at zero temperature. Phenomena arising from quantum fluctuations span a range of scales: from spontaneous emission and Lamb shifts of atoms~\cite{Milonni75, Lamb47}, intermolecular forces responsible for the stability of colloidal suspensions~\cite{PeterCE},  adhesive properties of gecko feet~\cite{gecko} and stiction in nano- and micro-mechanical systems~\cite{Serry98}, to potentially, density perturbations in the cosmic microwave background~\cite{mukhanov_2005}. Such quantum and thermal fluctuations of the EM field  can lead to forces between neutral objects, often addressed by different names depending on the geometry, separation and material properties of the interacting objects, such as van der Waals~\cite{van2004continuity}, London~\cite{London}, Casimir-Polder ~\cite{CP1948}, Casimir-Lifshitz ~\cite{Lifshitz}, or more generally Casimir~\cite{Casimir1948} forces, arise as a result of the interaction mediated between the fluctuating dipole moments that constitute two neutral bodies via the fluctuations of the EM field.

\begin{figure}
    \centering
\includegraphics[width = 0.55\textwidth]{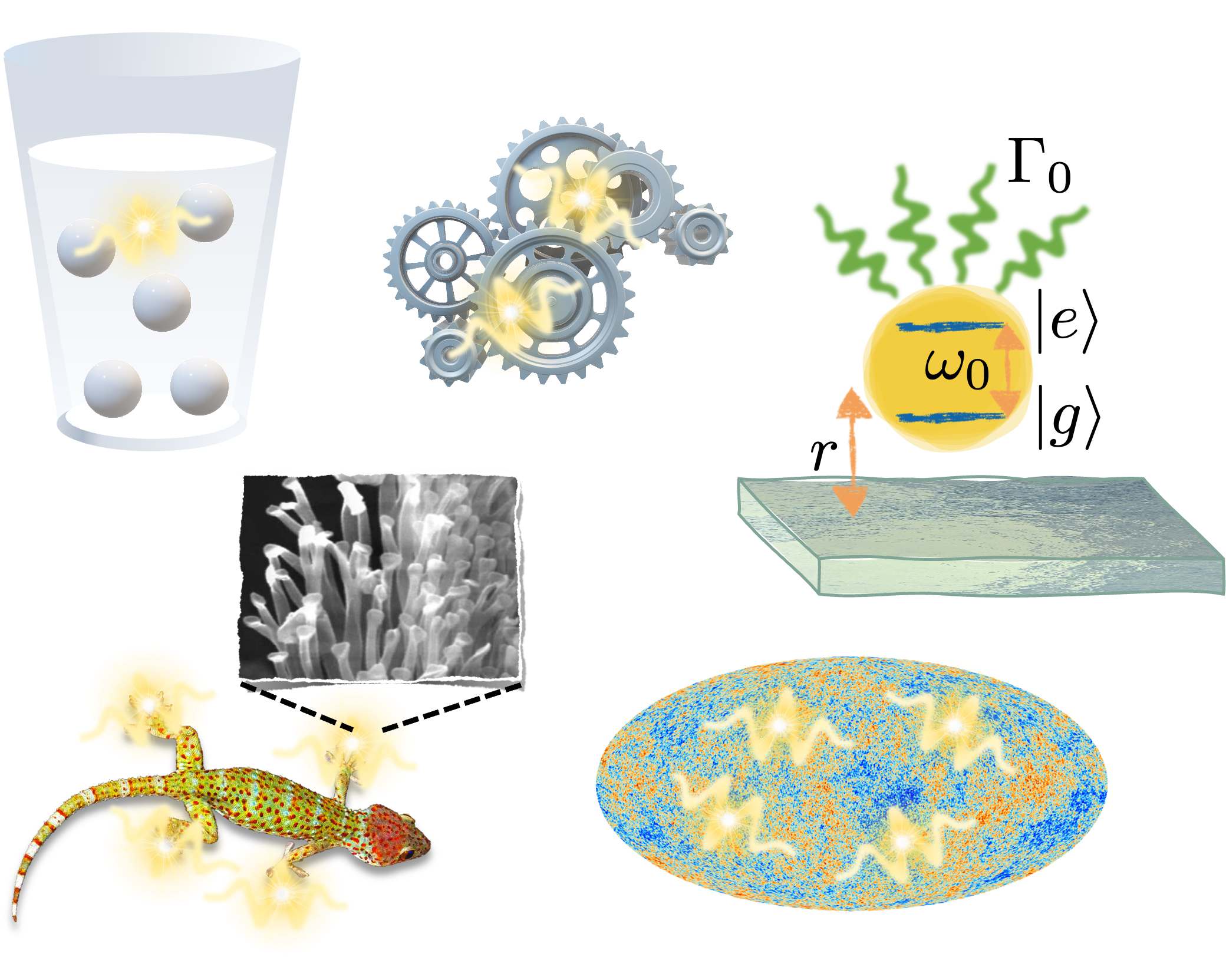}
\caption{Quantum fluctuation phenomena at various scales: Stability of colloids~\cite{PeterCE}, friction in  micro-electromechanical systems (MEMS)~\cite{Serry98}, van der Waals adhesion in gecko feet~\cite{gecko}, density perturbations in the Cosmic Microwave Background (CMB), and Casimir-Polder attraction of atoms near surfaces. Image credits: \cite{GeckoFeetImage, CMBimage}.}
    \label{fig:sch1}
\end{figure}

Fluctuation forces are typically studied between bodies that are classical. The development of nanoscale quantum systems opens an exciting new paradigm for exploring fluctuation forces between quantum systems that can be externally driven,  prepared in coherent superpositions or entangled states~\cite{Chang14, Fuchs18a, Fuchs18b, Sinha2020PRA, Fuchs_2018, CollCP18, Jones18, Behunin10}. This necessitates an Open Quantum System description of fluctuation forces, that we will review in the following. Such an approach enables one to treat  fluctuation-induced forces as well as the dissipation and decoherence effects  in the same framework.  To this end, we derive a master equation that describes  fluctuation-induced phenomena -- forces, dissipation and decoherence -- in a system of $N$ two-level atoms interacting with the quantum and thermal fluctuations of the EM field in the vicinity of a medium. While such an approach has been previously discussed in the context of analyzing  quantum fluctuation-induced forces~\cite{CollCP18}, here we take both the quantum and thermal fluctuations of the EM field into consideration as a  generalization of previous analyses.

\subsection{An Open Quantum System approach to Fluctuation Forces}

\subsubsection{Model and Hamiltonian}

\begin{figure}
    \centering
    \includegraphics[width = \textwidth]{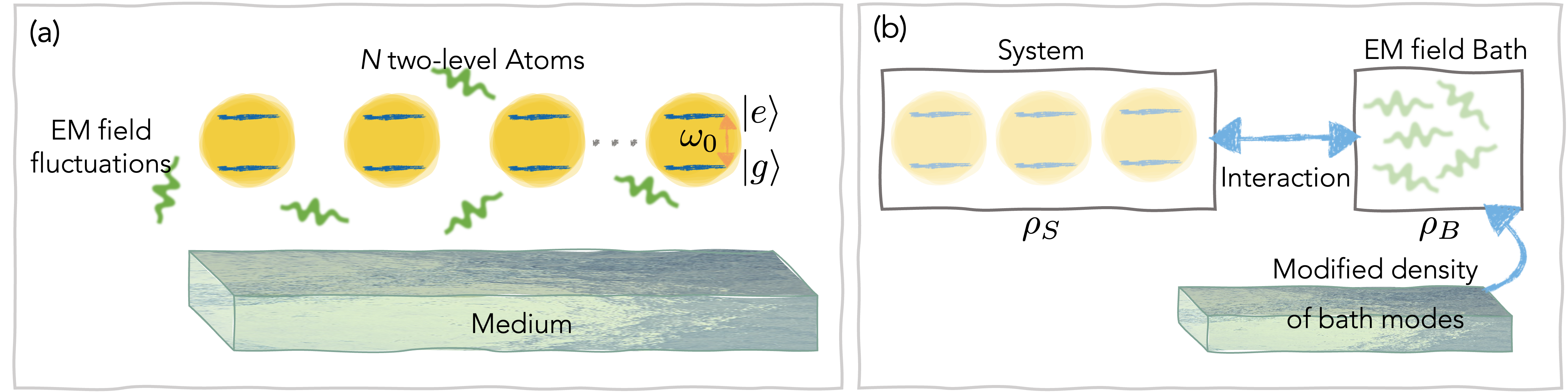}
    \caption{(a) Schematic representation of $N$ two-level atoms interacting with the quantum and thermal fluctuations of the EM field near a medium. Each atom has a resonant transition frequency $ \omega_0 $, with $\ket{g} $ and $ \ket{e}$ as the ground and excited levels.  (b) Open quantum system representation of the system in consideration. The collection of atoms is the system of interest, EM field is the bath, with its density of modes modified by the presence of the medium.}
    \label{Fig:Sch2}
\end{figure}

Consider a collection of $N$ identical two-level systems or atoms interacting with the fluctuations of the EM field, as shown in Fig.~\ref{Fig:Sch2}~(a). The  Hamiltonian describing the total system is,
\begin{equation}
H_\mr{tot} = H_S + H_B + H_\mr{int} \,, 
\end{equation}
where $H_S $ refers to the Hamiltonian for the atoms, 
\begin{equation}
H_S = \sum_{n = 1}^N \hbar \omega_0\, \sigma_n ^+ \sigma_n ^-
\end{equation}
with $\sigma_n ^+ $ and $\sigma_n ^- $ being the raising and lowering operators for the $n$th atom. Moreover, $\omega_0 $ is the resonance transition frequency.

The collection of atoms interacts with  the medium-assisted EM field that forms the bath, described by the Hamiltonian,
\begin{equation}
H_B = \int_0 ^\infty \dd \omega \int \dd^3 \mb{r}\sum_{\lambda = e,m}\mb{f}^\dagger_\lambda(\mb{r}, \omega)\mb{f}_\lambda(\mb{r}, \omega).
\end{equation}
The operators $ \mb{f}^{(\dagger)}_\lambda(\mb{r}, \omega)$ refer to the bosonic operators for the field in the presence of a medium as described in the macroscopic Quantum Electrodynamics (QED) formalism \cite{Buhmann1, Buhmann2, BuhmannRev, Scheel_Buhmann_2008}.   Physically one can understand the operators $ \mb{f}^{(\dagger)}_\lambda(\mb{r}, \omega)$ to represent the creation and annihilation operators for the polaritonic excitations of the field+matter system in the presence of media at frequency $\omega$, position $ \mb{r}$, with $ \lambda = e,m$ characterizing the quantum noise polarization $(e)$ or magnetization $(m)$. These operators obey the   canonical commutation relations,
\begin{equation} 
\sbkt{\mb{f}_\lambda\bkt{\mb{r}, \omega},\mb{f}_{\lambda'}^\dagger\bkt{\mb{r}', \omega'} }=\delta (\omega - \omega')\delta \bkt{\mb{r} - \mb{r}'}\delta_{\lambda\lambda'}\,.
\end{equation}

The system-bath interaction Hamiltonian is given by the electric-dipole  Hamiltonian~\cite{Milonni2019},
\begin{equation}
H_\mr{int} = \sum_n \mb{d}_n ( \sigma _n^+ +  \sigma _n ^ - )\cdot \mb{E}(\mb{r}_n),
\end{equation}
where $ \mb{d}_n$ is the dipole moment corresponding to the $n$th atom and $ \mb{E } \bkt{\mb{r}_n}$ is the quantized electric field at the position of the $n$th atom $ \mb{r}_n$, given by:
\begin{equation}
\mb{E}\bkt{\mb{r}_n} = \int_0 ^\infty \dd \omega \int \dd^3 \mb{r}\sum_\lambda \dbar{G}_\lambda \bkt{\mb{r}_n , \mb{r}, \omega}\cdot \mb{f}_\lambda \bkt{\mb{r}, \omega}+  \mb{f}^\dagger_\lambda \bkt{\mb{r}, \omega}\cdot \dbar{G}^\dagger_\lambda \bkt{\mb{r}_n , \mb{r}, \omega}.
\end{equation}
The coefficients $ \dbar G_\lambda (\mb{r}, \mb{r}' ,\omega)$ are proportional   to the Green's tensor  $ \dbar G\bkt{\mb{r}, \mb{r}',\omega}$ of the EM field,  accounting for the presence of any macroscopic media near the atoms, such that~\cite{Buhmann1, BuhmannRev, Scheel_Buhmann_2008}:
\begin{equation}
\label{eq:ggdag}
\sum_{\lambda = e,m}\int \dd^ 3 \mb{r}'\, \dbar{G}_\lambda \bkt{\mb{r}_1, \mb{r}', \omega}\cdot 
\dbar{G}^\dagger_\lambda \bkt{\mb{r}_2, \mb{r}', \omega} = \frac{\hbar \mu_0 }{\pi }\omega^2\, \mathfrak{Im}\left(\dbar {G} \bkt{\mb{r}_1 ,\mb{r}_2, \omega}\right),
\end{equation}
where $\mu_0$ is the vacuum magnetic permeability.

The Green's tensor $\dbar {G} \bkt{\mb{r}_1 ,\mb{r}_2, \omega} $ in the above equation represents the propagator for the EM field  in the presence of media, accounting for how a photon at frequency $ \omega$ propagates between positions $ \mb r_1 $ and $ \mb r_2$. The Green's tensor can be found as  the solution of the Helmholtz equation~\cite{Jackson},
\begin{equation}
\bkt{\nabla_1 \times \frac{1}{\mu(\mb{r}_1, \omega) }\nabla_1 \times - \frac{\omega^2 }{c^2}\epsilon\bkt{\mb {r}_1, \omega}}\dbar G (\mb{r}_1, \mb{r}_2 ,\omega) = \mb{\delta }\bkt{\mb{r}_1 - \mb{r_2}},
\end{equation}
where $\epsilon (\mb {r}, \omega)$ and $\mu(\mb{r}, \omega)$ represent the spatially dependent dielectric permittivity and magnetic permeability in the presence of a  medium. The Green's tensor encompasses the effects of surface geometry, optical and material properties of the medium, thereby modifying the density of states of the EM environment.

\subsubsection{Master equation}

We now continue to derive the open system dynamics of the atomic density matrix $ \rho _S $, by tracing out the medium-assisted quantized EM field. To this end, we first move to the interaction picture with respect to the  free Hamiltonian of the system and the bath $ H_0 = H_S + H_B$. As usual, we define the interaction Hamiltonian in the interaction picture as $ \tilde H_ \mr{int}\equiv \e{-i H_0 t/\hbar }\, H_\mr{int}\, \e{i H_0 t/\hbar}$,
\begin{equation}
\begin{split}
\tilde H_\mr{int}\bkt{t} &=  \sum_n\int_0 ^\infty \dd \omega \int \dd^3 \mb{r}\sum_\lambda\bkt{ \sigma _n^+ e^{i \omega_0 t} +  \sigma _n ^ -e^{-i \omega_0 t} }\\
&\times \sbkt{ \mb{d}_n \cdot  \dbar{G}_\lambda \bkt{\mb{r}_n , \mb{r}, \omega}\cdot \mb{f}_\lambda \bkt{\mb{r}, \omega}e^{-i \omega t}+  \mb{f}^\dagger_\lambda \bkt{\mb{r}, \omega}\cdot \dbar{G}^\dagger_\lambda \bkt{\mb{r}_n , \mb{r}, \omega}\cdot \mb{d}_n e^{i \omega t}}.
\end{split}
\end{equation}
The second-order Born-Markov master equation that describes the evolution of the atomic density matrix is thus given by \cite{BPbook},
\begin{equation}
\label{eq:mast}
\begin{split}
\der{\rho_S }{t} &= - \frac{1}{\hbar^2}\, \ptr{B}{\int _0 ^\infty \dd\tau\, \sbkt{\tilde H_\mr{int}(t) , \sbkt{\tilde H_ \mr{int} (t - \tau),\rho_S \otimes \rho_B }}}\\
&= \underbrace{ - \frac{1}{\hbar^2}\, \ptr{B}{\int _0 ^\infty \dd\tau\, \tilde H _\mr{int} \bkt{t}\tilde H _\mr{int} \bkt{t- \tau }\rho_S \otimes\rho_B} }_\mr{(I)}\\
&\qquad \underbrace{- \frac{1}{\hbar^2} \ptr{B}{\int _0 ^\infty \dd\tau\, \rho_S \otimes\rho_B \tilde H _\mr{int} \bkt{t- \tau }\tilde H _\mr{int} \bkt{t}}}_{\mr{(II)}}\\
&\qquad\underbrace{+ \frac{1}{\hbar^2}\, \ptr{B}{\int _0 ^\infty \dd\tau\, \tilde H _\mr{int} \bkt{t}\rho_S \otimes\rho_B \tilde H _\mr{int} \bkt{t- \tau }}}_{\mr{(III)}}\\
&\qquad\underbrace{+ \frac{1}{\hbar^2}\, \ptr{B}{\int _0 ^\infty \dd\tau\,\tilde H _\mr{int} \bkt{t- \tau }  \rho_S \otimes\rho_B \tilde H _\mr{int} \bkt{t}}}_{\mr{(IV)}}
\end{split}
\end{equation}
In making the Born approximation in the above master equation we assume that the state of the EM field bath is always in thermal equilibrium at some given temperature $T$, such that the bath equilibration time scale is much faster than the system evolution. The thermal density matrix of the bath is  given by $\rho_B = \e{- H_B /(k_B T)}/Z$, with $ Z $ as the partition function.

We can simplify each of the terms in the above master equation \eqref{eq:mast}, starting with (I),
\begin{equation}
\label{eq:11}
\begin{split}
\mr{(I)}&=  - \frac{1}{\hbar^2}\, \ptr{B}{\int _0 ^\infty \dd\tau\, \tilde H _\mr{int} \bkt{t}\tilde H _\mr{int} \bkt{t- \tau }\rho_S \otimes\rho_B} \\
&=  - \frac{1}{\hbar^2}\,\mrm{tr}_\mrm{B}\bigg\{\int_0 ^\infty \dd\tau \sum_{n,m}\int_0 ^\infty \dd \omega\dd \omega' \int \dd^3 \mb{r}\dd^3\mb{r}'\sum_{\lambda,\lambda'}\bkt{ \sigma _n^+ e^{i \omega_0 t} +  \sigma _n ^ -e^{-i \omega_0 t} }\non\\
&\quad\times\sbkt{ \mb{d}_n \cdot  \dbar{G}_\lambda\cdot \mb{f}_\lambda\bkt{\mb{r}, \omega } e^{-i \omega t}+  \mb{f}_\lambda^\dagger\bkt{\mb{r}, \omega }\cdot \dbar{G}^\dagger_\lambda\cdot \mb{d}_n e^{i \omega t}}\bkt{ \sigma _m^+ e^{i \omega_0 (t - \tau)} +  \sigma _m ^ -e^{-i \omega_0 (t - \tau)}} \non\\
&\quad\times\sbkt{ \mb{d}_m \cdot  \dbar{G}_{\lambda'}\cdot \mb{f}_{\lambda'}\bkt{\mb{r}', \omega '}e^{-i \omega'(t-\tau)}+  \mb{f}^\dagger_{\lambda'}\bkt{\mb{r}', \omega '}\cdot \dbar{G}^\dagger_{\lambda'}\cdot \mb{d}_m e^{i \omega' (t-\tau)}}\rho_S \otimes \rho_B\bigg\},
\end{split}
\end{equation}
where we have defined the shorthand notation $\dbar G_\lambda^{(\dagger)}\equiv \dbar G^{(\dagger)}_{\lambda}\bkt{\mb{r_n}, \mb{r}, \omega} $ and  $\dbar G_{\lambda'}^{(\dagger)}\equiv \dbar G^{(\dagger)}_{\lambda'}\bkt{\mb{r_m}, \mb{r}', \omega'} $. To take the trace over the field, we note that,
\begin{equation}
\begin{split}
\ptr{B}{\mb{f}_1^\dagger \mb f_2 \rho_B} = & n_\mr{th}(\omega_1) \delta (\omega_1 - \omega_2)\delta(\mb{r}_1 - \mb{r}_2)\delta_{\lambda\lambda'} \\
\ptr{B}{\mb{f}_1 \mb f_2^\dagger \rho_B} = & \bkt{n_\mr{th}(\omega_1) + 1} \delta (\omega_1 - \omega_2)\delta(\mb{r}_1 - \mb{r}_2)\delta_{\lambda\lambda'}\\
\ptr{B}{\mb{f}_1 \mb f_2 \rho_B} = & \ptr{B}{\mb{f}_1 ^\dagger\mb f_2^\dagger \rho_B} =0
\end{split}
\end{equation}
where $ n_\mr{th}(\omega) = 1/(\e{\hbar \omega /(k_B T)} -1)$ is the average thermal occupation number in mode $ \omega$. Tracing over the bath and using the fluctuation-dissipation relation for the Green's tensor~\eqref{eq:ggdag},  Eq.~\eqref{eq:11} can be thus simplified to,
\eqn{&\mr{(I)} = -\frac{\mu_0 }{\pi \hbar} \int_0 ^\infty \dd\tau \sum_{n,m} \int_0 ^\infty\dd\omega ~\omega^2\mb{d}_n \cdot  \im \dbar{G}(\mb{r}_n,\mb{r}_m, \omega)\cdot \mb{d}_m\non\\
&\sbkt{ \bkt{n_\mr{th} (\omega) + 1} e^{-i \omega \tau}+  n_\mr{th} (\omega) e^{i \omega\tau}}\bkt{\hat \sigma _n^+\hat \sigma _m ^ -e^{i \omega_0  \tau}  + \hat \sigma _n ^ - \hat \sigma _m^+ e^{-i \omega_0  \tau} } \rho_S.
}
Performing the integral over $ \tau$ and $ \omega$, we get\footnote{The integration over $ \tau $ involves the use of the indentiy $ \int_0 ^\infty \dd\tau e^{i X \tau } = \pi \delta (X) + i \mc{P}\frac{1}{X}$. The dirac-delta function contributes to the dissipative coefficients $\Gamma_{nm}$ and the principal value part to the dispersive energy shifts $\Delta \omega_{nm}$. The  dispersive shifts are then obtained via performing contour integrals over  $\omega $, the details of the derivation can be found in \cite{Buhmann2}.}:

\eqn{\label{eq:me1}\mr{(I)} =   \sum_{m,n} &\sbkt{\cbkt{- \frac{\Gamma_{nm}}{2} \bkt{ n_\mr{th} (\omega_0 )+ 1} + i \bkt{\Delta\omega_{nm}^\mr{OR}-  \Delta\omega_{nm}^\mr{R}\bkt{ n_\mr{th}(\omega_0 ) + 1}  }}\hat \sigma_n^+ \hat \sigma_m ^-\right.\non\\
&\left. +   \cbkt{- \frac{\Gamma_{nm}}{2}n_\mr{th}(\omega_0 )  - i \bkt{\Delta\omega_{nm}^\mr{OR} + \Delta \omega _{nm}^\mr{R}n_\mr{th}(\omega_0 ) }}\hat \sigma_n^- \hat \sigma_m ^+ }\rho_S,
}
where we have defined

\eqn{
\Delta \omega_{nm}^\mr{R} \equiv & \frac{\mu_0 }{\hbar} \omega_0 ^2  \mb{d}_n \cdot \mathfrak{Re} \dbar G \bkt{\mb{r}_n, \mb{r}_m, \omega_0 }\cdot \mb{d}_m\\
\Gamma_{nm}\equiv&\frac{2\mu_0 \omega_0^2}{\hbar}    \mb{d}_n \cdot \mathfrak{Im} \dbar G \bkt{\mb{r}_n, \mb{r}_m, \omega_0 }\cdot \mb{d}_m\\
\Delta \omega_{nm}^{\mr{OR}}\equiv&\frac{2\mu_0 \omega_0 k_B T}{\hbar^2} {\sum_{p = 0}^\infty }' \frac{\omega_p ^2 }{\omega_p ^2 + \omega_0 ^2} \mb{d}_n \cdot  \dbar G \bkt{\mb{r}_n, \mb{r}_m,  i \omega_p }\cdot \mb{d}_m,
}
Here $ \Delta \omega_{nm}^\mr{R}$ represents the resonant dipole-dipole interaction between atoms $n$ and $m$ mediated by  fluctuations of the EM field in the presence of a medium. Similarly,  $ \Gamma_{nm}$  denote  the dissipative interaction coefficients between atoms $ n$ and $m$, induced by thermal and quantum fluctuations.  We note that the resonant contribution  to the total Casimir-Polder potential $ \Delta\omega _{nm}^\mr{R}$, as well as its dissipative counterpart  $\Gamma_{nm}$, depend on the real and imaginary part of the response function  of the environment at the resonant frequency of the atoms ($ \re\dbar{G} \bkt{\mb{r}_n,\mb{r}_m, \omega_0}$ and $ \im\dbar{G} \bkt{\mb{r}_n,\mb{r}_m, \omega_0}$). The off-resonant interaction coefficient $\Delta \omega_{nm}^{\mr{OR}}$ comprises  contribution from the fluctuations at all frequencies, with  
$ \omega_p =  p\frac{2 \pi k_B T  }{\hbar} $  $( p \in \mathbb{Z}^+)$, representing the Matsubara frequencies~\cite{BuhmannScheel2008, Buhmann2}. The prime on the summation indicates that the $ p = 0$ term comes with a factor of $ 1/2$.

We can now simplify  the remaining terms in the master equation to obtain:

\eqn{\label{eq:me2}\mr{(II)} =& \sum_{m,n}\rho_S \sbkt{ \cbkt{- \frac{\Gamma_{nm}}{2}\bkt{n _\mr{th}(\omega_0 )  + 1} -i \bkt{\Delta \omega_{nm}^\mr{OR} -\Delta \omega_{nm}^\mr{R} \bkt{n _\mr{th}(\omega_0 )  + 1}   }}\hat \sigma_n ^+  \hat \sigma_m ^-\right.\non\\
&\left.+ \cbkt{- \frac{\Gamma_{nm}}{2}n _\mr{th}\bkt{\omega_0 } + i\bkt{ \Delta \omega_{nm}^\mr{OR} +\Delta \omega_{nm}^\mr{R}n _\mr{th}\bkt{\omega_0 } }}\hat \sigma_n ^-  \hat \sigma_m ^+}\\
\label{eq:me34}
\mr{(III)} =  &\mr{(IV)} = \sum_{m,n}\frac{\Gamma_{nm} }{2}  \bkt{n _\mr{th}(\omega_0 )  + 1} \hat \sigma_m^-  \rho_S\hat  \sigma_n^+ +\frac{\Gamma_{nm}}{2} n _\mr{th}(\omega_0 )    \hat \sigma_n^-  \rho_S \hat \sigma_m^+.
}

Adding together Eqs.~\eqref{eq:me1}, \eqref{eq:me2} and \eqref{eq:me34}, we obtain the full master equation as:

\eqn{\label{eq:ME}
\der{\rho_S }{t} = - \frac{i}{\hbar} \sbkt{H_\mr{eff}, \rho_S} + \mc{L}_\mr{eff}\sbkt{\rho_S}}

where $H_\mr {eff}$ is the medium-assisted effective Hamiltonian that describes the energy shifts or fluctuation forces on the system,  given by:

\eqn{\label{Eq:ham}
H_\mr{eff} =&\sum_{n=1}^N \hbar \sbkt{\Delta \omega_{nn}^\mr{OR} - \Delta \omega_{nn}^\mr{R}\bkt{n_\mr{th}\bkt{\omega_0 } + 1}} \hat\sigma^+_{n} \hat\sigma^-_{n} - \hbar \sbkt{\Delta \omega_{nn}^\mr{OR} \right.\non\\
&\left.+ \Delta \omega_{nn}^\mr{R}n_\mr{th}\bkt{\omega_0 }}\hat \sigma^-_{n}\hat \sigma^+_{n}- \sum_{n\neq m } \hbar \bkt{\Delta \omega_{nm}^\mr{R} \bkt{2n_\mr{th}\bkt{\omega_0 } + 1}} \hat\sigma^+_{n} \hat\sigma^-_{m} 
}
 The medium-assisted Liouvillian  $ \mc{L}_\mr{eff}$ accounts for the dissipation and decoherence of the system, and is given by the Lindblad superoperator:
\eqn{\mc{L}_\mr{eff}\sbkt{\rho_S} =\sum_{n,m = 1}^N &-\frac{\Gamma_{nm}}{2} \bkt{n_\mr{th}(\omega_0 ) + 1}\sbkt{\hat \sigma^+_{n}\hat\sigma^-_{m}\rho_S + \rho_S\hat\sigma^+_{n}\hat\sigma^-_{m} -2 \hat\sigma^-_{n}\rho_S\hat\sigma^+_{m}}\non\\
& -\frac{\Gamma_{nm}}{2}n_\mr{th}(\omega_0 ) \sbkt{\hat \sigma^-_{n}\hat\sigma^+_{m}\rho_S + \rho_S\hat\sigma^-_{n}\hat\sigma^+_{m} -2 \hat\sigma^+_{n}\rho_S\hat\sigma^-_{m}}, 
}
where the first term denotes dissipation and decoherence due to both thermal and quantum fluctuations. The second term represents absorption of thermal radiation by the atomic system, and vanishes at $T\rightarrow0$.

The Hamiltonian $H_\mr{eff}$ accounts for the fluctuation-induced energy shifts or forces in the presence of external media for a collection of atoms in arbitrary states. The fluctuation-induced force between the atoms and the medium can be found as the spatial derivative of the medium-modified energy of the atomic system,
\begin{equation}
    F = \der{}{z}\,\ptr{S}{\rho_S H_\mr{eff}},
\end{equation}
where $ z$ is the coordinate along which the atoms and medium are separated from each other.
    
\begin{figure}
    \centering
    \includegraphics[width = 0.8\textwidth]{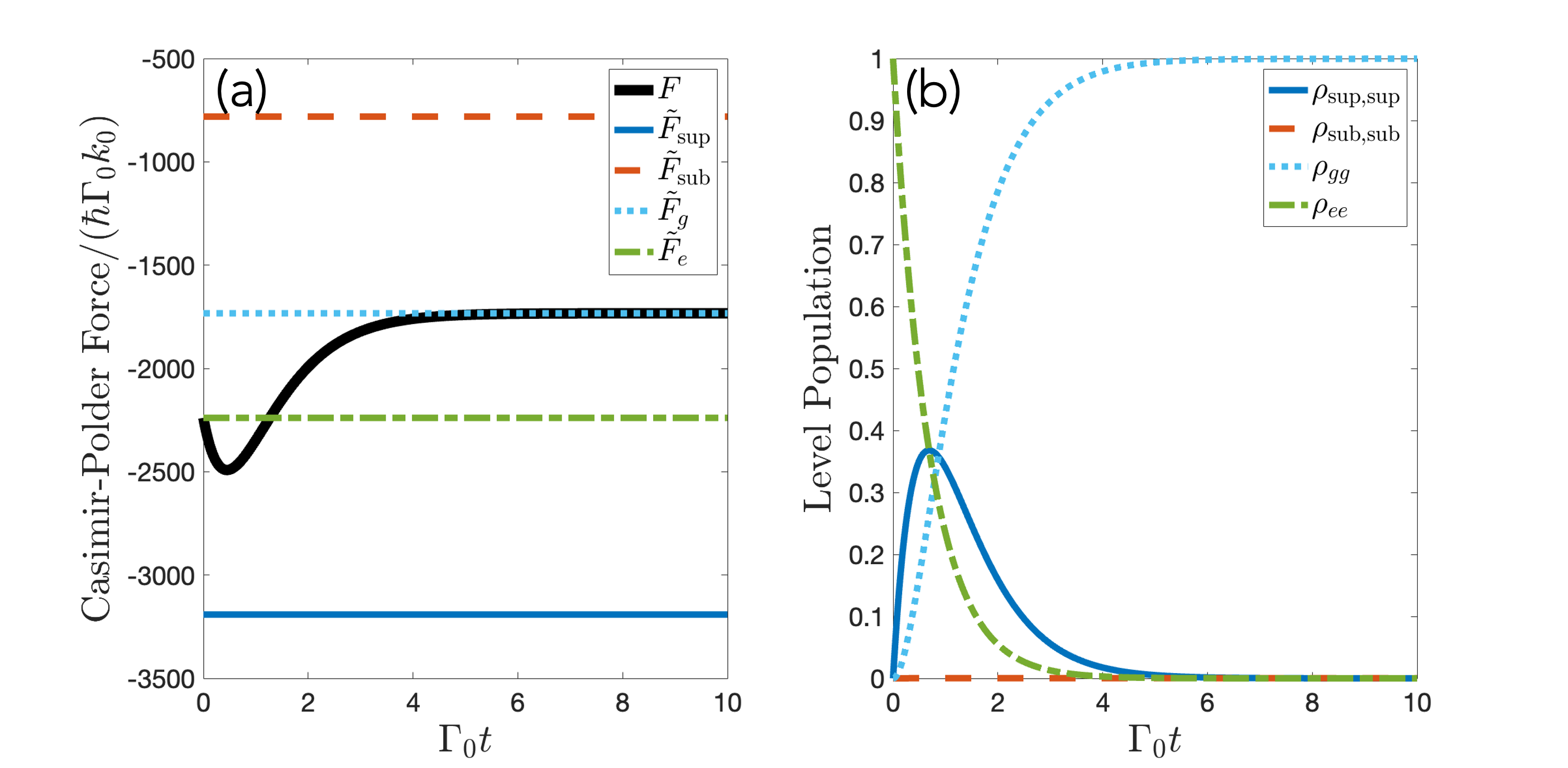}
    \caption{Fluctuation-induced force on a system of two atoms near a silica surface. The atoms are initially prepared in excited states $\ket{ee}$, as can be seen from the green dash-dotted curve in (b). As the atomic system interacts with the EM field bath, the atoms get de-excited, eventually going to the ground state as depicted by the blue dotted curve in (b). The total  fluctuation force on the atoms is represented  by the black solid curve in (a). There is an enhancement in the force as the atomic population in the superradiant state increases.  }
    \label{Fig:coll}
\end{figure}

\subsubsection{Instructive example: two atoms near a surface}

As an example, we can analyze the fluctuation-forces on two atoms prepared in collective states near a surface, as shown in Fig.~\ref{Fig:coll}. It can be seen that starting with the initial state $\rho_S = \ket{ee}\bra{ee}$,  the dynamics of the two atoms is  such that their joint density matrix   can be written as
\begin{equation}
\rho_S = \rho_{ee}\ket{ee}\bra{ee} + \rho_\mr{sup, sup}\ket{\psi_+}\bra{\psi_+}+ \rho_\mr{sub, sub}\ket{\psi_-}\bra{\psi_-} + \rho_{gg}\ket{gg}\bra{gg}, 
\end{equation}
where $ \ket{\psi_\pm} = \bkt{\ket{eg}\pm \ket{ge}}/\sqrt{2}$ refer to the Dicke super- and subradiant states, that are known to exhibit enhancement and suppression in spontaneous emission~\cite{Dicke, Gross}.  The blue solid curve $(\tilde F_\mr{sup})$ and the red dashed curve $(\tilde F_\mr{sub})$ in  Fig.~\ref{Fig:coll}(a) show that these states exhibit a similar enhancement and suppression in the fluctuation-induced force~\cite{CollCP18, Fuchs_2018, Jones18}. The master equation  approach thus allows one to monitor the non-equilibrium dynamics of fluctuation-induced forces on the  collective atomic density matrix in the presence of correlations.

\subsection{Remarks}
We have introduced a master equation description of quantum and thermal fluctuation-induced forces on a system of $N $ two-level atoms near a medium.   The  master equation~\eqref{eq:ME} is sufficiently general to describe the non-equilibrium dynamics of  the system prepared in general quantum state near an arbitrary macroscopic body and includes all fluctuation phenomena -- forces/energy shifts, dissipation and decoherence. Our approach combines the macroscopic QED formalism~\cite{Buhmann1, Buhmann2, BuhmannRev, Scheel_Buhmann_2008} with an Open Quantum System framework~\cite{BPbook}, allowing one to analyze fluctuation-induced forces and dissipative dynamics in quantum systems on the same footing.

There is an interesting parallel between the QED and thermodynamic perspectives on fluctuation-induced forces. In the QED framework,  fluctuation-induced forces  arise from the sum over interactions between the fluctuating dipole moments that constitute matter, mediated via virtual and thermal photons of the quantized EM field. Equivalently, these forces can be seen as originating via the difference in the free energy of a photon gas with and without the presence of external boundaries~\cite{Mehra67, Brown69,  Scharnhorst87, Revzen97,  Mitter2000, Bezerra2002, Intravaia_2008}. In addition to fluctuation-induced forces, the framework of fluctuational QED has been successfully applied to study radiative heat transfer between bodies in close proximity~\cite{Polder71, Pendry99, Biehs11, Kruger11, Messina2011, Messina2011b, Bimonte17, BiehsRMP}. Moreover, the concepts of non-equilibrium thermodynamics have been used to demonstrate quantum friction in atoms moving near surfaces~\cite{Intravaia14, Reiche20, Reiche20b}.

The natural question arises whether thermodynamic considerations can further constrain and/or characterize the properties of quantum  and thermal fluctuation-induced phenomena. 

\section{Second law of thermodynamics for fluctuations}

Some of the most important cornerstones of quantum thermodynamics are statements of the second law that are collectively known as \emph{fluctuation theorems}. The first versions were derived for purely classical dynamics~\cite{jarzynski97,crooks1999entropy}, which are already regarded as the most significant results in the modern thermodynamics~\cite{Ortiz2011}. Many important principles naturally follow, such as the arrow of time~\cite{jarzynski2011equalities} and linear response theory~\cite{andrieux2008quantum,andrieux2009fluctuation}. To better explore the laws of thermodynamics at the nanoscale, reformulating the second law of thermodynamics from the fundamentals of quantum physics is, thus, of particular significance. 

\subsection{Two-Time Measurement Scheme}
\label{sec:TPM}

In the energy representation, the two central quantities of thermodynamics are heat and work \cite{Callen1998}. In contrast to state functions, though, these two quantities are described by non-exact differentials. In other words, heat and work depend on the specific path a thermodynamics system takes in state space. This is consistent with classical mechanics, where work is defined as a functional of a force along a trajectory.

Due to Heisenberg's uncertainty relations, classical trajectories are somewhat ill-defined in quantum dynamics. Hence, how to define work and heat in quantum systems becomes significantly more involved \cite{DeffnerBook19}. In complete analogy to classical thermodynamics, ``meaningful'' notions of quantum thermodynamic work are motivated by experimental accessibility.

Arguably, the most prominent approach has been called two-time measurement (TTM) (or two-point measurement) scheme~\cite{Tasaki00,Kurchan01,Morikuni17,Jarzynski15,Gardas2018,kiely2023entropy,Campisi11,mazzola2013measuring,dorner2013extracting}. In this paradigm, thermodynamic work is identified as the difference of the outcomes of energy measurements preformed at the beginning and the end of a process. If the system is isolated, i.e., if it evolves under unitary dynamics
\begin{equation}
\partial_{t} U_{t}=-i/\hbar\, H_{t}U_{t}\quad\text{with}\quad U_0=\id\,,
\end{equation}
any difference in energy has to be identified with work. In other words, isolated quantum systems do not exchange energy with the environment. 

\begin{figure}
\centering
\includegraphics[width=.95\columnwidth]{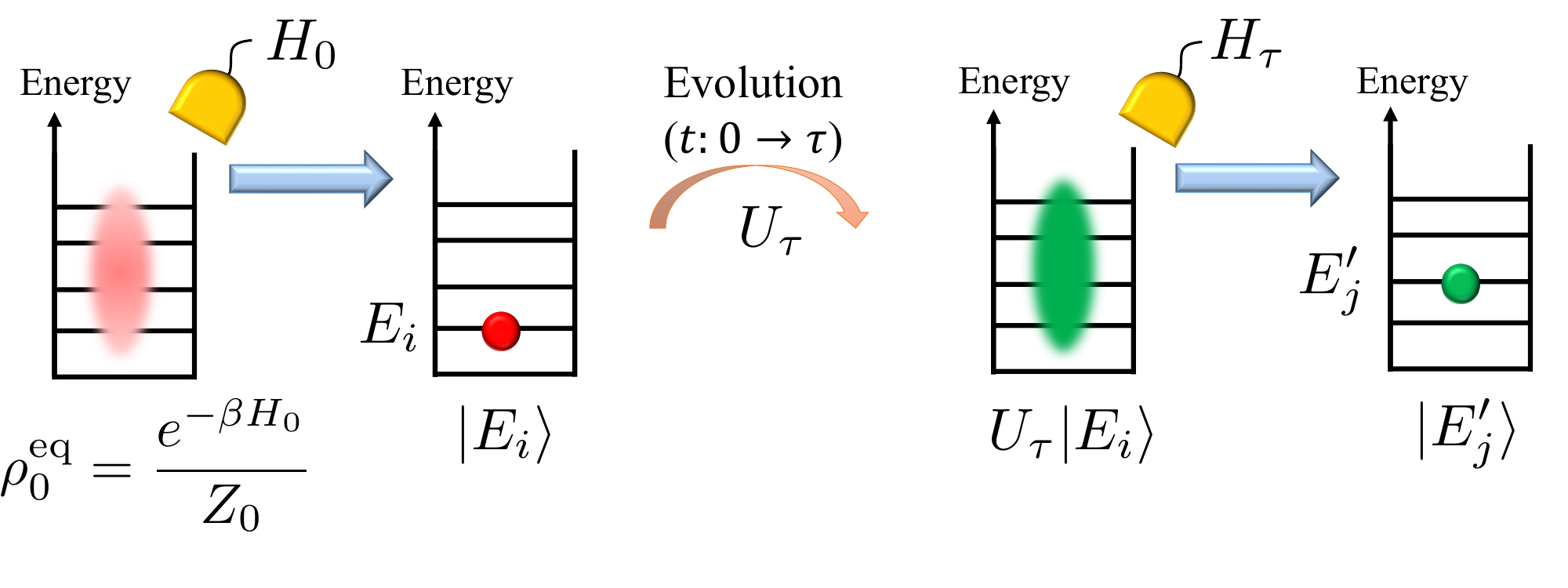}
\caption{In the TTM scheme, the quantum system is initially prepared in a Gibbs state $\rho_{0}^{\eq}=\e{-\beta H_0}/Z_0$. A first energy measurement is performed, which projects the state into an eigenstate of $H_0$, $\ket{E_i}$ with the measurement outcome $E_i$. Then, the system is let to evolve under the unitary $U_{\tau}$, before a the second energy measurement $H_{\tau}$ is performed.}
\label{fig:TPM}
\end{figure}

Schematically the TTM scheme is illustrated in Fig.~\ref{fig:TPM}. 
Consider a system initially prepared in a Gibbs state with inverse temperature $\beta$ and initial Hamiltonian $H_0$, $\rho_{0}^{\text{eq}} = \e{-\beta H_0}/Z_0$, where as before $Z_0\equiv\tr{\e{-\beta H_0}}$ is the standard partition function. At $t=0$ a first projective measurement of the energy is performed, which projects the system into an energy eigenstate $\{E_i,\ket{E_i}\}$ of $H_0$. The corresponding eigenvalue $E_i$ is recorded, and the system  is let to evolve under the unitary $U_t$ from $t=0$ to $t=\tau$. At $t=\tau$ a second measurement of the energy is performed, which now projects the system into an eigenstate$\{E_j',\ket{E_j'}\}$ of $H_{\tau}$.

The first law of thermodynamics then dictates that the energy difference is the work is the thermodynamic work,
\begin{equation}
    W_{i\to j}=E_j'-E_i\,.
\end{equation}
The corresponding probability density function characterizing the fluctuations of the work reads
\begin{equation}
    P_F(W)\equiv\sum_{ij}\delta(W-E_j'+E_i)\,\abs{\bramatket{E_j'}{U_{\tau}}{E_i}}^2\,\frac{e^{-\beta E_i}}{Z_0}\,,
\end{equation}
where $\abs{\bramatket{E_j'}{U_{\tau}}{E_i}}^2$ is the transition probability from $\ket{E_i}$ to $\ket{E_j'}$. Note that the such defined work distribution is consistent with thermodynamics, which we observe from
\begin{equation}
\begin{split}
    \ave{W}_{P} &= \int dW P_F(W) W=\sum_{ij} \frac{e^{-\beta E_i}}{Z_0}\abs{\bramatket{E_j'}{U_{\tau}}{E_i}}^2 (E_j'-E_i)\\
    &=\tr{H_{\tau}U_{\tau}\rho_{0}^{\text{eq}}U_{\tau}\ad}-\tr{H_{0}\rho_{0}^{\text{eq}}}=\avg{H_\tau}-\avg{H_0}\equiv \ave{W}\,.
\end{split}
\end{equation}

In standard thermodynamics \cite{Callen1998}, reversible processes are ideal in the sense that they dissipate the least amount of energy into the environment. Hence, it is often also interesting to consider the ``reverse'' quantum process. To this end, consider the same quantum system, but now initially prepared in a Gibbs state with respect to the final Hamiltonian, $H_\tau$, and which evolves under the ``backward'' dynamics $U_t^\dagger$. The corresponding work distribution reads
\begin{equation}
    P_B(-W)\equiv\sum_{ij}\delta(-W-E_i+E_j')\,\abs{\bramatket{E_i}{U_{\tau}\ad}{E_j'}}^2\frac{e^{-\beta E_j'}}{Z_{\tau}}\,,
\label{eq:backward}
\end{equation}
where $Z_{\tau}\equiv \tr{\e{-\beta H_{\tau}}}$ is the standard partition function of $H_{\tau}$. 

Interestingly, the work distributions for forward and backward processes are closely related. We have,
\begin{equation}
\begin{split}
P_F(W) &=\sum_{ij}\frac{e^{-\beta E_i}}{Z_0}\abs{\bramatket{E_j'}{U_{\tau}}{E_i}}^2\delta(W-E_j'+E_i)\\
&=\frac{Z_{\tau}}{Z_0}\sum_{ij}\frac{e^{-\beta (E_j'-W)}}{Z_{\tau}}\abs{\bramatket{E_j'}{U_{\tau}}{E_i}}^2\delta(W-E_j'+E_i)\\
&=\frac{Z_{\tau}}{Z_0}\left(\sum_{ij}\frac{e^{-\beta E_j'}}{Z_{\tau}}\abs{\bramatket{E_i}{U_{\tau}\ad}{E_j'}}^2\delta(-W-E_i+E_j')\right)e^{\beta W}\,.
\end{split}
\end{equation}
Hence, we obtain the Quantum Crooks Fluctuation Theorem~\cite{Tasaki00,Kurchan01}
\begin{equation}
    \frac{P_F(W)}{P_B(-W)} =e^{\beta(W-\Delta F)}\,,
\label{eq:FT}
\end{equation}
where $\Delta F\equiv-\beta^{-1}\ln(Z_{\tau}/Z_{0})$ is the equilibrium free energy difference. 

The Quantum Crooks Fluctuation Theorem \eqref{eq:FT} is a so-called detailed fluctuation theorem. The corresponding integral version, the Quantum  Jarzynski Equality \cite{Campisi11,Talkner16} is simply obtained from
\begin{equation}
    \ave{e^{-\beta W}}\equiv \int dW e^{-\beta W}P_F(W) = e^{-\beta \Delta F}\,.
\end{equation}

It is important to recognize that the fluctuation theorems permit to determine the free energy difference from an average over nonequilibrium fluctuations of the thermodynamic work. This is a stronger statement then the standard maximum work theorem, which provides only an inequality. In fact, employing Jensen's inequality, the Jarzynski equality directly gives
\begin{equation}
    \ave{W}\geq \Delta F\,.
\end{equation}

Finally, it has been noted that the TTM scheme has a good correspondence with the classical scenario~\cite{jarzynski2015quantum,Zhu16}, which makes it a conceptually sound starting point for quantum thermodynamic considerations.

\subsubsection{Symmetry relation for the characteristic function}

The fluctuation theorems can be written in a more convenient form by considering the characteristic functions of the work distributions~\cite{Campisi11}. Let $C_F(u)$ and $C_B(u)$ be the characteristic functions of $P_F(W)$ and $P_B(-W)$, respectively
\begin{equation}
\begin{split}
C_F(u) \equiv \int dW P_F(W) e^{iuW}\quad \text{and}\quad C_B(u) \equiv \int dW P_B(-W) e^{-iuW}\,,
\end{split}
\end{equation}
which can be written as
\begin{equation}
\begin{split}
    C_F(u) &= \sum_{ij}\frac{e^{iu E_j'+i(-u+i\beta)E_i}}{Z_0}\abs{\bramatket{E_j'}{U_{\tau}}{E_i}}^2\\
    C_B(u)& = \sum_{ij}\frac{e^{i(-u+i\beta)E_j'+iu E_i}}{Z_{\tau}}\abs{\bramatket{E_i}{U_{\tau}\ad}{E_j'}}^2\,.
\end{split}
\end{equation}
Therefore, we immediately have
\begin{equation}
    \frac{C_F(u)}{C_B(-u+i\beta)}=e^{-\beta \Delta F}\,,
\label{eq:symmetric2}
\end{equation}
which expresses the Quantum Crooks Fluctuation theorem~\eqref{eq:FT} as a symmetry relation for the characteristic functions. In the following, Sec.~\ref{sec:QC}, we will demonstrate that $C_F(u)$ and $C_B(-u+i\beta)$ can be computed by using the single-qubit interferometry algorithm on a quantum computer.

\subsection{Quantum circuits for verifying quantum fluctuation theorem}
\label{sec:QC}

The quantum fluctuation theorem based on the TTM scheme has been experimentally studied using various platforms, such as trapped ions~\cite{Huber2008,Smith2018,An15}, nuclear magnetic resonance~\cite{Tiago14}, defect spins~\cite{hernandez2020experimental,hernandez2021experimental}, superconducting qubit system~\cite{Zhenxing18,hahn2022verification}, and even the D-Wave machine \cite{Gardas2018SR}. 

In the remainder of the present exposition, we focus on how to implement the symmetry relation \eqref{eq:symmetric2} on quantum computers. To this end, we re-write the expressions for the characteristic function as
\begin{equation}
C_F(u) =\tr{U^\dagger_\tau e^{iuH_{\tau}} U_\tau e^{-iuH_0}\rho_0^{\eq}}\,.
\label{eq:CharacForward2}
\end{equation}
and
\begin{equation}
C_B(u) =\tr{U_\tau e^{iuH_0} U^\dagger_\tau e^{-iuH_{\tau}}\rho_{\tau}^{\eq}}\,,
\label{eq:CharacBackward2}
\end{equation}
where $\rho_{\tau}^{\eq}=\e{-\beta H_{\tau}}/Z_{\tau}$ denotes the Gibbs state defined by the final Hamiltonian $H_{\tau}$. We will now see, that the expressions of Eqs.~\eqref{eq:CharacForward2} and \eqref{eq:CharacBackward2} are particularly useful as they can be computed using single-qubit interferometry,
which was experimentally implemented in Refs.~\cite{mazzola2013measuring, dorner2013extracting}. 

\subsubsection{General Idea}

\begin{figure}
\center\mbox{
\Qcircuit @C=1em @R=1.2em {
\lstick{\ket{0}}& \gate{\mathbf{H}}& \ctrl{0} \qwx[1] & \qw& \meter~~~~~~~~\text{$\sigma_x$ and $\sigma_y$}  \\
\lstick{\rho}&\qw & \gate{U} & \qw&\qw \\
}
}
\caption{Quantum circuit of the single-qubit interferometry}
\label{fig:circuit0}
\end{figure}
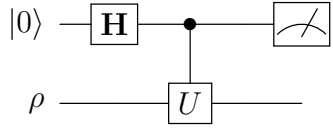

Consider the single-qubit interferometry quantum circuit, depicted in Fig.~\ref{fig:circuit0}. Here, $\mathbf{H}$ is the Hadamard gate, which can be expressed in the logical basis as
\begin{equation} 
\mathbf{H}\equiv \frac{1}{\sqrt{2}}
\begin{pmatrix}
1&1\\
1&-1
\end{pmatrix}
\end{equation}
and $U$ is a unitary operator. We further define $\ket{0}\equiv \begin{pmatrix}1&0\end{pmatrix}^T$ and $\ket{1}\equiv\begin{pmatrix}0&1\end{pmatrix}^T$. Choosing the input state to be
\begin{equation} 
\rho_{\text{in}}=\dya{0}\otimes\rho=\begin{pmatrix}\rho&0\\0&0\end{pmatrix}\,,
\end{equation}
it is easy to see that the output state is given by 
\begin{equation}
\rho_{\text{out}} =\frac{1}{2}
\begin{pmatrix}
\rho&\rho U\ad\\
U\rho&U\rho U\ad
\end{pmatrix}\,.
\end{equation}

As before, $\sigma_x,\sigma_y,\sigma_z$ denote the Pauli's matrices 
\begin{align}
\sigma_x=\begin{pmatrix}0&1\\1&0\end{pmatrix},~
\sigma_y=\begin{pmatrix}0&-i\\i&0\end{pmatrix},~
\quad\text{and}\quad \sigma_z=\begin{pmatrix}1&0\\0&-1\end{pmatrix}.
\end{align}
Thus, we have
\begin{equation}
\ave{\sigma_x} =\tr{(\sigma_x\otimes\id)\rho_{\text{out}}}=\mf{Re}\left(\tr{\rho U}\right)\,,
\end{equation}
and 
\begin{equation}
\ave{\sigma_y} =\tr{(\sigma_y\otimes\id)\rho_{\text{out}}}=\mf{Im}\left(\tr{\rho U}\right)\,.
\end{equation}
Consequently, we obtain $\tr{\rho U}$ from the quantum circuit in Fig.~\ref{fig:circuit0} as  
\begin{align}
\tr{\rho U} = \ave{\sigma_x}+i\ave{\sigma_y}\,.
\end{align}

\subsubsection{Quantum circuits for computing characteristic functions}

We will now see that the characteristic functions can be computed by employing the same single-qubit interferometry. The quantum circuit for computing $C_F(u)$ is depicted in Fig.~\ref{fig:circuitKF}, and Fig.~\ref{fig:circuitKB} shows the the quantum circuit for computing $C_B(u)$.  
\begin{figure}[H]
\center\mbox{
\Qcircuit @C=1em @R=1.2em {
\lstick{\ket{0}}& \gate{\mathbf{H}}& \ctrl{0} \qwx[1] &
\qw & \ctrl{0} \qwx[1] & \meter&~~~~~~~~\text{$\sigma_x$ and $\sigma_y$} \\
\lstick{\rho_0^{\eq}}&\qw & \gate{e^{-i u H_0}} & \gate{U_{\tau}}& \gate{e^{iu H_{\tau}}} &\qw \\
}
}
\caption{Quantum circuit for computing $C_F(u)$}
\label{fig:circuitKF}
\end{figure}
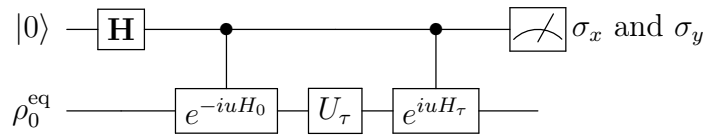
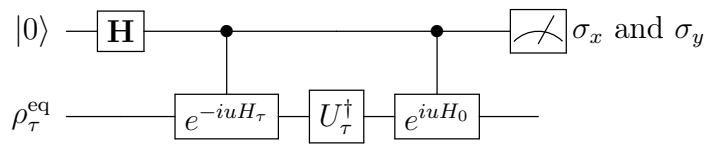
\begin{figure}[H]
\center\mbox{
\Qcircuit @C=1em @R=1.2em {
\lstick{\ket{0}}& \gate{\mathbf{H}}& \ctrl{0} \qwx[1] &\qw & \ctrl{0} \qwx[1] & \meter&~~~~~~~~\text{$\sigma_x$ and $\sigma_y$} \\
\lstick{\rho_{\tau}^{\eq}}&\qw & \gate{e^{-i u H_{\tau}}} & \gate{U_{\tau}\ad}& \gate{e^{iu H_0}} &\qw \\
}
}
\caption{Quantum circuit for computing $C_B(u)$.}
\label{fig:circuitKB}
\end{figure}

For the symmetry relation \eqref{eq:symmetric2}, we are interested in $C_B(-u+i\beta)=\tr{U_{\tau}e^{-\beta H_0} e^{-i u H_0}U_{\tau}\ad e^{\beta H_{\tau}}e^{i u H_{\tau}}\rho_{\tau}^{\eq}}$. A minor technical difficulty arises from the fact that $\e{\beta H_{\tau}}$ and $\e{-\beta H_0}$, are not unitary operators, which means that they cannot be represented by a quantum gate. Hence, to compute $C_B(-u+i\beta)$, we need employ the method developed in Ref.~\cite{maeda2023detailed}. First,  $\e{\beta H_{\tau}}$ and $\e{-\beta H_0}$  are decomposed into a Pauli sequence $\{\sigma_{\ell}\}_{\ell=1}^{d^2}$, where $\sigma_1\equiv \id$~\footnote{For example, for a two-qubit system $(d=4)$, the Pauli sequence is the set of operators $\{\id, \sigma_x\otimes \id, \sigma_y\otimes\id, \sigma_z\otimes\id, \id\otimes \sigma_x, \id\otimes \sigma_y, \id\otimes \sigma_z, \sigma_x\otimes \sigma_x, \sigma_x\otimes \sigma_y, \sigma_x\otimes \sigma_z, \sigma_y\otimes \sigma_x, \sigma_y\otimes \sigma_y, \sigma_y\otimes \sigma_z, \sigma_z\otimes \sigma_x, \sigma_z\otimes \sigma_y, \sigma_z\otimes \sigma_z \}$, which has $4^2=16$ elements, and these elements are the bases constructing any $4\times 4$ matrix. Usually, for a $d\times d$ matrix, the Pauli sequence has $d^2$ elements.}.
Then, we can write
\begin{equation}
e^{-\beta H_0} = \sum_{\ell=1}^{d^2}\alpha_{\ell}^{(0)}\sigma_{\ell}\quad \text{and}\quad e^{\beta H_{\tau}}= \sum_{\ell=1}^{d^2}\alpha_{\ell}^{(\tau)}\sigma_{\ell}\,,
\end{equation}
where we have
\begin{equation}
\alpha_{\ell}^{(0)}= \frac{1}{d}\, \tr{e^{-\beta H_0}\sigma_{\ell}}\quad\text{and}\quad\alpha_{\ell}^{(\tau)}=\frac{1}{d}\, \tr{e^{\beta H_{\tau}}\sigma_{\ell}}\,.
\end{equation}
These parameters can be regarded as known quantities since we have full knowledge of $H_0$ and $H_{\tau}$.

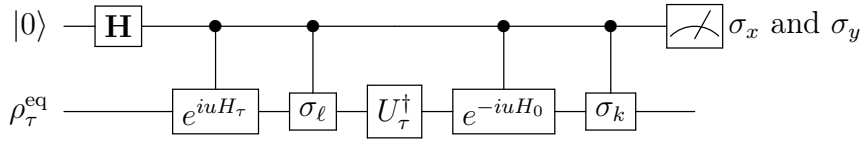
\begin{figure}
\center\mbox{
\Qcircuit @C=1em @R=1.2em {
\lstick{\ket{0}}& \gate{\mathbf{H}}& \ctrl{0} \qwx[1] &
\ctrl{0} \qwx[1] & \qw & \ctrl{0} \qwx[1] & \ctrl{0} \qwx[1] &  \meter&~~~~~~~~\text{$\sigma_x$ and $\sigma_y$} \\
\lstick{\rho_{\tau}^{\eq}}&\qw & \gate{e^{iu H_{\tau}}}&\gate{\sigma_{\ell}} & \gate{U_{\tau}\ad}& \gate{e^{-iuH_0}}& \gate{\sigma_k} &\qw \\
}
}
\caption{Quantum circuit for computing $F_{k\ell}$ to construct $C_B(-u+i\beta)$.}
\label{fig:circuitKBimaginary}
\end{figure}

With this, we can write
\begin{align}
C_{B}(-u+i\beta) = \sum_{k\ell}\alpha_{k}^{(0)}\alpha_{\ell}^{(\tau)} \tr{U_{\tau}\sigma_{k} e^{-iu H_0}U_{\tau}\ad \sigma_{\ell}e^{iu H_{\tau}}\rho_{\tau}^{\eq}}.
\label{eq:CimaginaryDecompose}
\end{align}
Defining
\begin{align}
F_{k\ell} \equiv \tr{U_{\tau}\sigma_{k} e^{-iu H_0}U_{\tau}\ad \sigma_{\ell}e^{iu H_{\tau}}\rho_{\tau}^{\eq}}\,,
\end{align}
we can construct the quantum circuit depicted in Fig.~\ref{fig:circuitKBimaginary}
to compute $F_{k\ell}$. In conclusion, the characteristic functions can be directly determined with the depicted quantum circuits.

However, we would be remiss if we forwent a few remarks on the shortcomings of the TTM scheme. It has even been argued that TTM scheme is thermodynamically inconsistent since the projective energy measurements inevitably destroy quantum coherences~\cite{Marti17}. To overcome this issue, recent works proposed alternative paradigms, such as dynamic Bayesian networks~\cite{Micadei20}, the Maggenau-Hill quasiprobability~\cite{Levy2020}, and the one-time measurement (OTM) scheme~\cite{Deffner16,Sone20a,Sone21b,Beyer2020,sone2023jarzynskilike}. Particularly, in the OTM scheme, the distribution of changes in internal energy is constructed by considering the expectation value of the energy conditioned on the initial energy measurement outcomes. This formalism avoids the second projective measurement and, therefore, the thermodynamic contribution of quantum coherence or the correlations generated by the dynamics  are naturally contained in the formalism \cite{Sone2021entropy}. However, we have shown that the OTM scheme is nothing but the TTM scheme for non-demolition measurements, which can also be implemented with quantum circuits \cite{Maeda2023}.

\subsection{Fluctuation theorems for open quantum dynamics}

The natural question arises how the fluctuation theorem can be generalized to open system dynamics described by master equations such as \eqref{eq:ME} in an experimentally meaningful way. While this is still a somewhat open question, it has been suggested that also for open systems TTM is a conceptually sound approach \cite{Campisi09,Deffner2011PRL,Kafri12,rastegin2013non,Rastegin14,goold2021fluctuation,Touil2021PRXQ}. However, also other approaches such as OTM \cite{Sone20a} or alternative notions in phase space \cite{Deffner2011EPL,Deffner2013EPL} have been considered. 

\section{Concluding remarks}

It is a core tenet of statistical mechanics that in macroscopic systems fluctuations are irrelevant. Mathematically, this is underpinned by the law of large numbers. However, at the nanoscale fluctuations dominate and hence any thermodynamically consistent description will have to focus on the nature, characteristics, and consequences of fluctuations. In this chapter of the \emph{Encyclopedia of Mathematical Physics} we have outlined two conceptually related, but typically separately considered areas of current research: fluctuation forces and fluctuation theorems. The confluence of these two topics holds significant potential in terms of building a non-equilibrium quantum thermodynamics-based description of fluctuation phenomena in QED, with possible applications in  near-field radiative heat transfer,  design of nanoscale quantum systems and  fundamental connections between thermodynamics and fluctuational QED that remain to be elucidated. 

\begin{flushleft}
    \textbf{Acknowledgements}
\end{flushleft}
A.S is supported by the startup package from University of Massachusetts Boston. This work was supported  by the National Science Foundation under Grant No. PHY-2309341 (K.S.) and No. PHY-1748958 (K.S.), and by the John Templeton Foundation under Award No. 62422 (K.S. and S.D.).

\bibliographystyle{elsarticle-num-names} 
\bibliography{encyclo_bib}

\end{document}